\documentclass[fleqn,usenatbib]{mnras}

\usepackage{newtxtext,newtxmath}
\usepackage{float}

\usepackage{placeins}
\usepackage{adjustbox} 
\usepackage{graphicx}
\usepackage{float}
\usepackage{lscape}
\usepackage{wrapfig}
\usepackage{epstopdf}
\usepackage{subfigure}
\usepackage{epstopdf}
\DeclareGraphicsExtensions{.ps}
\usepackage{float}
\usepackage{booktabs}
\usepackage{adjustbox} 
\usepackage{placeins}

\usepackage[T1]{fontenc}

\DeclareRobustCommand{\VAN}[3]{#2}
\let\VANthebibliography\thebibliography
\def\thebibliography{\DeclareRobustCommand{\VAN}[3]{##3}\VANthebibliography}

\usepackage{graphicx}	
\usepackage{amsmath}
\usepackage{amssymb}	

\title{The first X-ray spectrum of the HMXB XTE J1855$-$026 during the compact object eclipse.\footnote{Based on the observation obtained by {\it XMM-Newton} observatory.}}

\author[]{
Sanjurjo-Ferr\'{i}n, G.$^{1}$
Torrej\'on, J.M.$^{1}$
Rodes-Roca, J.J.$^{1}$
\\
$^{1}$Instituto Universitario de F\'isca Aplicada a las Ciencias y las Tecnolog\'ias, Universidad de Alicante, 03690 Alicante, Spain\\
}

\date{Accepted XXX. Received YYY; in original form ZZZ}

\pubyear{2020}

\begin{document}
\label{firstpage}
\pagerange{\pageref{firstpage}--\pageref{lastpage}}
\maketitle

\begin{abstract}
We present the first {\it XMM-Newton} observation of the classical supergiant high-mass X-ray binary XTE J1855$-$026 taken entirely during the eclipse of the neutron star (NS), covering the orbital phases $\phi= 0.00-0.11$. The analysis of the data allows us to a) compare with the parameters obtained during the existing pre eclipse observation and b) explore the back illuminated stellar wind of the B0I type donor. The black body component, used to describe the soft excess during pre eclipse, is not observed during eclipse. It must be then produced near the NS or along the donor-NS line. The $0.3-10$ keV luminosity during eclipse ($\sim 10^{34}$ erg s$^{-1}$) is 70 times lower than pre eclipse. The intensity of the Fe K$\alpha$ line, in the average eclipse spectrum, is $\sim 7.4$ times lower than the one measured during pre eclipse. Since K$\alpha$ photons can not be resonantly scattered in the wind, the vast majority of Fe K$\alpha$ emission must come from distances within $1R_{*}$ from the NS. The eclipse spectrum is successfully modelled through the addition of two photoionized plasmas, one with low ionization ($\log\xi_{\rm 1,cold}=0.36$) and high emission measure ($EM_{\rm 1,cold}\approx 3\times 10^{59}$ cm$^{-3}$) and another with high ionization ($\log\xi_{\rm 2,hot}=3.7$) and low emission measure ($EM_{\rm hot}\approx 2\times 10^{56}$ cm$^{-3}$). Assuming that the cold and hot gas phases are the clumps and the interclump medium of the stellar wind, respectively, and a clump volume filling factor of $\approx [0.04-0.05]$, typical for massive stars, a density contrast between clumps and the interclump medium of $n_{\rm c}/n_{\rm i}\approx 180$ is deduced, in agreement with theoretical expectations and optical-UV observations of massive star winds.

\end{abstract}

\begin{keywords}
Stars:massive -- X-rays: binaries -- X-rays: individual: XTE J1855$-$026
\end{keywords}



\section{Introduction}

Supergiant X-ray binaries (SGXBs) are the high mass X-ray binaries (HMXBs) where a compact object (a neutron star -NS- or a black hole) orbits an evolved  massive star (the companion) in its supergiant phase, accreting matter from its powerful stellar wind. HMXBs have been a prime target since the dawn of X-ray astronomy \citep[see][for a recent review]{2019NewAR..8601546K}.  Interest in these systems have being revamped in the last years for two reasons. First, they are the natural progenitors of the double degenerate binaries whose coalescence produces the gravitational waves predicted by the General Theory of Relativity and now finally detected \citep[][]{PhysRevLett.116.061102,2019IAUS..346....1V}. To characterise the physical properties of the parent population is of paramount importance. Second, they are prime laboratories to study the stellar winds in massive stars \citep[][]{2017SSRv..212...59M}. Massive stars ($M_{\rm i}>8M_{\odot}$) are among the main drivers of the evolution of star clusters and galaxies. Their powerful stellar winds and their final supernova explosions inject large amounts of matter and mechanical energy into their environments, thus enriching the interstellar medium and further triggering star formation. Yet the structure and properties of massive star winds are still poorly known.

The accretion of matter from stellar wind onto an NS powers strong X-ray radiation which, in turn, illuminates nearby wind regions. This radiation excites transitions in the stellar wind, producing emission lines of different elements. These lines intensities change relative to continuum with the orbital phase, been specially enhanced during the eclipse, when the direct continuum produced by the neutron star is blocked by the optical counterpart \citep[][]{2015ApJ...810..102T, 2019ApJS..243...29A, 2021MNRAS.501.5646M}. Thus, eclipsing systems with supergiant companions are particularly well suited to study the irradiated stellar wind. The stellar wind properties ($v_{\infty}$, $\dot{M}$, $\rho(r)$) and the binary system characteristics ($R_{*}$, $a$) combine to influence the observed X-ray spectrum (most notably, through the ionization parameter $\xi=L_{\rm X}/n(r_{\rm X})r^{2}_{\rm X}$) and these change, among other things, with the donor's spectral type. A continuum of types would, thus, be desirable but, unfortunately, there are only a handful of such systems. Characterizing them all is, thus, very important. 

\begin{figure}
\includegraphics[trim={0cm 8cm 0cm 1cm},width=1\columnwidth]{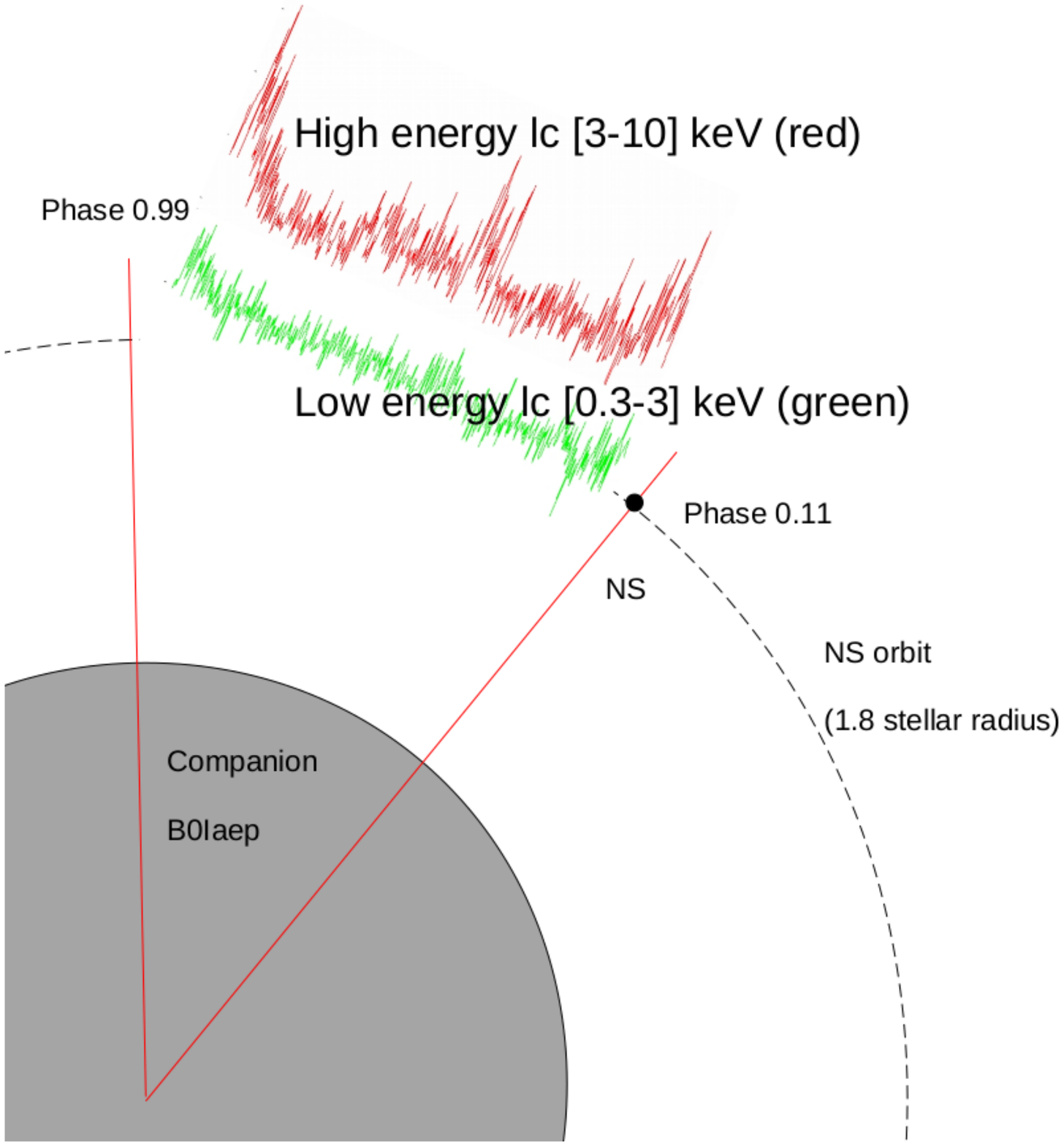}
\caption{Pole-on sketch of the system and orbital phases covered by the {\it XMM-Newton} observation, using the ephemerides of \citet{falanga}. The donor star radius, and the orbit are to scale. }
\label{systemsk}
\end{figure}

The X-ray source, XTE J1855$-$026, was discovered by the \emph{Rossi X-ray Timing Explorer} (\emph{RXTE}) satellite \citep{Corbet_1999}. The source contains a NS showing a $\simeq 361$ s X-ray pulse, orbiting the companion every $\sim$ 6 d. Through the analysis of the eclipse duration \citet{Corbet_2002} suggest a massive companion with a radius corresponding to a B0I donor. This is further supported through the Spectral Energy Distribution (SED) fitting \citep[][]{2013ApJ...764..185C}, as well as through direct optical spectrum fitting, which refines the spectral type to B0Iaep \citep[][]{gonzalezgalan2016fundamental}. The distance to XTE J1855$-$026, derived from the European Space Agency (ESA) mission
{\it Gaia}\footnote{(\url{https://www.cosmos.esa.int/gaia})} is $7.4 \pm 0.8$ kpc, using the combined parallax measure and the source's G-band magnitude and BP-RP colour \citep[][]{2021AJ....161..147B}.  In Table \ref{parameters} we compile the system parameters relevant for this work. In Fig. \ref{systemsk} we show a sketch of the system where the orbit and the donor radius are to scale.

\citet{2018JApA...39....7D} present the only low-resolution CCD X-ray spectral analysis using \textit{Suzaku} data. This observation was performed entirely out of eclipse, just prior to ingress. In this paper we present the analysis of the first observation of XTE J1855$-$026, taken entirely during eclipse, using the X-ray Multi-Mirror Mission ({\it XMM-Newton}) space observatory. This data are used to analyse the emission line spectrum with unprecedented detail. 



\begin{table}
\centering
\begin{tabular}{lcc}
\hline\hline\\
\multicolumn{3}{c}{Companion}\\
\midrule
    MK type & B0Iaep &  1,2\\
    $M_{\rm opt} $ &   21 $M_\odot$ &  \\
    
    $R_\ast$ & 22 $R_\odot$ &  \\
          &       &  \\
\multicolumn{3}{c}{Neutron star}\\
\midrule
     $M_{\rm NS}$ &   1.4 $M_\odot$  &  1\\
    Spin period &   360.7 s &  1 \\
     $\dot P/ P$ & -12(13) $\times  10^{-6}$ yr$^{-1}$ & 3\\  
            &       &  \\
\multicolumn{3}{c}{System}\\ 
\midrule
    Orbital period  & 6.07415(8) d  &  3\\
    $i$ & $ ~ 71(2){\rm deg}$ & 3 \\
    Eccentricity  & 0.04(2)  & 1 \\
    Orbital radius &   1.8~ $R_\ast$ & Deduced\\
    Orbital velocity &   330 km s$^{-1}$ & Deduced\\
     Distance &  $7.4 \pm 0.8$ kpc  & 4,5\\
    $T_{0}$ (MJD) & 52 704.009(17)  &  3  \\
  \hline
\end{tabular}
\caption{Properties of XTE J1855$-$026 system. (1) \citet{Corbet_2002}, (2) \citet{gonzalezgalan2016fundamental}, (3) \citet{falanga}, (4) \citet{2021AJ....161..147B}, (5) \citet{vo:gedr3dist_main} }
\label{parameters}
\end{table}

\section{Observation and analysis.}
The {\it XMM-Newton} spacecraft carries three high throughput X-ray telescopes and one optical monitor. The \textit{European Photon Imaging Camera} (EPIC) focal plane instruments, pn, MOS1 and MOS2, provide broad band coverage ($E\sim[0.3-10]$ keV) with a moderate spectral resolution ($E/\Delta E\sim 20-50$). The \textit{Reflection Grating Spectrometer} (RGS) provides high resolution spectra ($E/\Delta E\sim 150-800$) over a limited spectral range $E\sim[0.3-2.1]$ keV). In Table \ref{xmmobs} we present the observation log. 

\begin{table}
\caption {\emph{XMM-Newton} Observation log. }
\centering
\begin{adjustbox}{max width=\columnwidth}
\begin{tabular}{cccc}
\hline\hline
Observation ID & Date & Orbital phase & Duration\\
  &    &  &  (ks)\\
\midrule
   0844630101 &  2020/3/21   &   $0.99-0.11$ &  60  \\
  \hline

\end{tabular}
\end{adjustbox}
\label{xmmobs}
\end{table}

\text

The observation was carried out using medium filters for the three EPIC focal plane instruments MOS1, MOS2 and pn. The three cameras were operating in large window mode. The data were first processed through the pipeline chains and filtered. For MOS1 and MOS2, only events with a pattern between 0 and 12  were considered, filtered through \#XMMEA EM. For pn, we kept events with \texttt{flag=0} and a pattern between 0 and 4 \citep{2001A&A...365L..27T}. The chosen extraction region was a circle centered in the brightest point of the source. The background selected was an annulus around the extraction region. We checked whether the observations were affected by pile-up, using the \texttt{epatplot} task, with negative results.

The spectra were produced with a spectral bin size of 6 and analyzed and modelled with the \textit{Interactive Spectral Interpretation System} (\textsc{isis}) package \citep{2002hrxs.confE..17H}\footnote{maintained by MIT at \url{https://space.mit.edu/cxc/isis/}}. The data from the three cameras, MOS1, MOS2 and pn, was finally combined using the task \texttt{epicspeccombine} for the analysis. The energy range used for spectral fitting was $0.35-10$ keV. The errors were obtained with the \texttt{fit$\_$pars} and the \texttt{conf} tasks, provided by \textsc{isis}, for a 90$\%$ confidence level. The emission lines were identified thanks to the \textsc{atomdb}\footnote{http://www.atomdb.org/} data base and the \textit{X-Ray Data Booklet} \citep{thompson2001x}.

The lightcurve timing analysis was performed by combining the light curves from the three cameras (MOS1, MOS2 and pn) using the task \texttt{lcmath}.
The photon arrival times were transformed to the solar system barycentre. 

The observation took place entirely during the eclipse of the X-ray source, as the 99\% of the X-ray flux is occulted between phases 0.92 and 0.097 \citep{falanga}. Even when the last flare is out of this range, it is only by $\phi_{\rm orb}=1.3^{\circ}$ and a return to a low number of counts is observed at the end of the section.


\section{Results}

\subsection{X-ray lightcurve}

\begin{figure}
\centering
\subfigure{\includegraphics[trim={1cm 6cm 0cm 2cm},width=1\columnwidth]{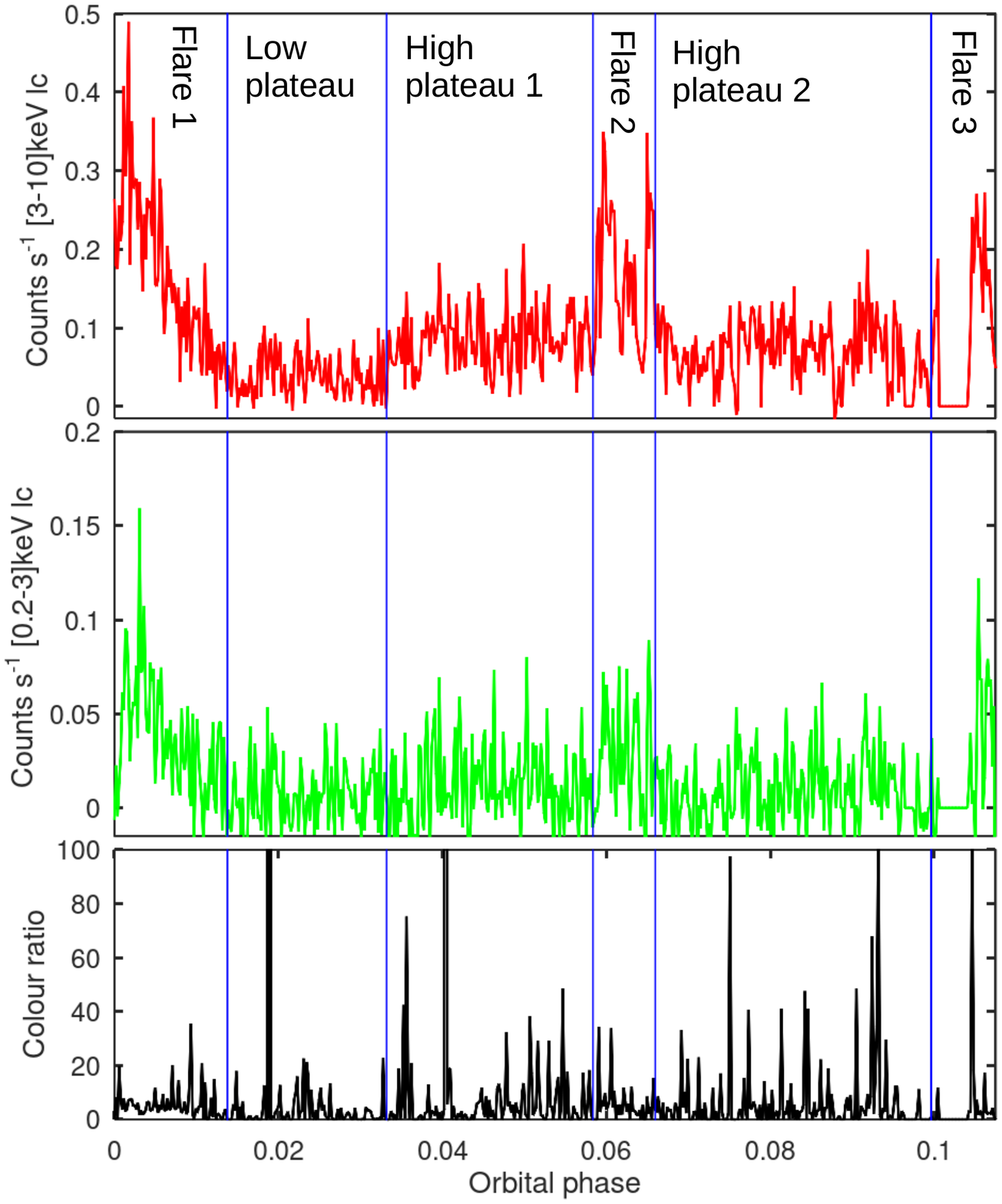}}
\caption{\textit{XMM-Newton} EPIC lightcurves of XTE J1855$-$026 for the $3-10$ keV (red, upper panel), $0.2-3$ keV (green, middle panel) energy ranges along with the color ratio $CR=(3-10)/(0.2-3)$ (black, lower panel). The divisions show the six different sections in which the observation was divided, according to the source flux. The time bin is 150 s}.
\label{lcurve}
\end{figure}

The lightcurve, produced combining data from the three EPIC cameras, is shown in Fig. \ref{lcurve} for the $3-10$ keV (red) and $0.2-3$ keV (green) energy ranges.  The color ratio $CR=(3-10)/(0.2-3)$ (black) is also plotted. In general, it looks stable. This is expected during the X-ray eclipse. However, some variability is still observed. Consequently, the light curve was further divided in six intervals: three plateaus (two high and one low) and three flares. The count rate of each interval is presented in Table \ref{counts_interval}. These intervals will be used for separate spectral analysis in the next section.

\begin{table}
\centering		
\caption{Weighted average count rate per interval. (Fig. \ref{lcurve}).}
\begin{adjustbox}{max width=\columnwidth}		
\begin{tabular}{rr}		
\toprule		
& Weighted mean (cts s$^{-1}$)\\
&$\times10^{-2}$\\
\midrule		
Flare 1	&	$17 \pm 10$\\
Low plateau	&	$4\pm 3$\\
High plateau 1	&	$8 \pm 4$\\
Flare 2	&	$17\pm 9$\\
High plateau 2	&	$ 8 \pm 3$\\
Flare 3	&	$15 \pm 7$\\
\bottomrule		
\end{tabular}		
\end{adjustbox}		
\label{counts_interval}		
\end{table}	

\begin{figure}
\subfigure{\includegraphics[trim={1cm 5cm 0cm 1cm},width=1\columnwidth]{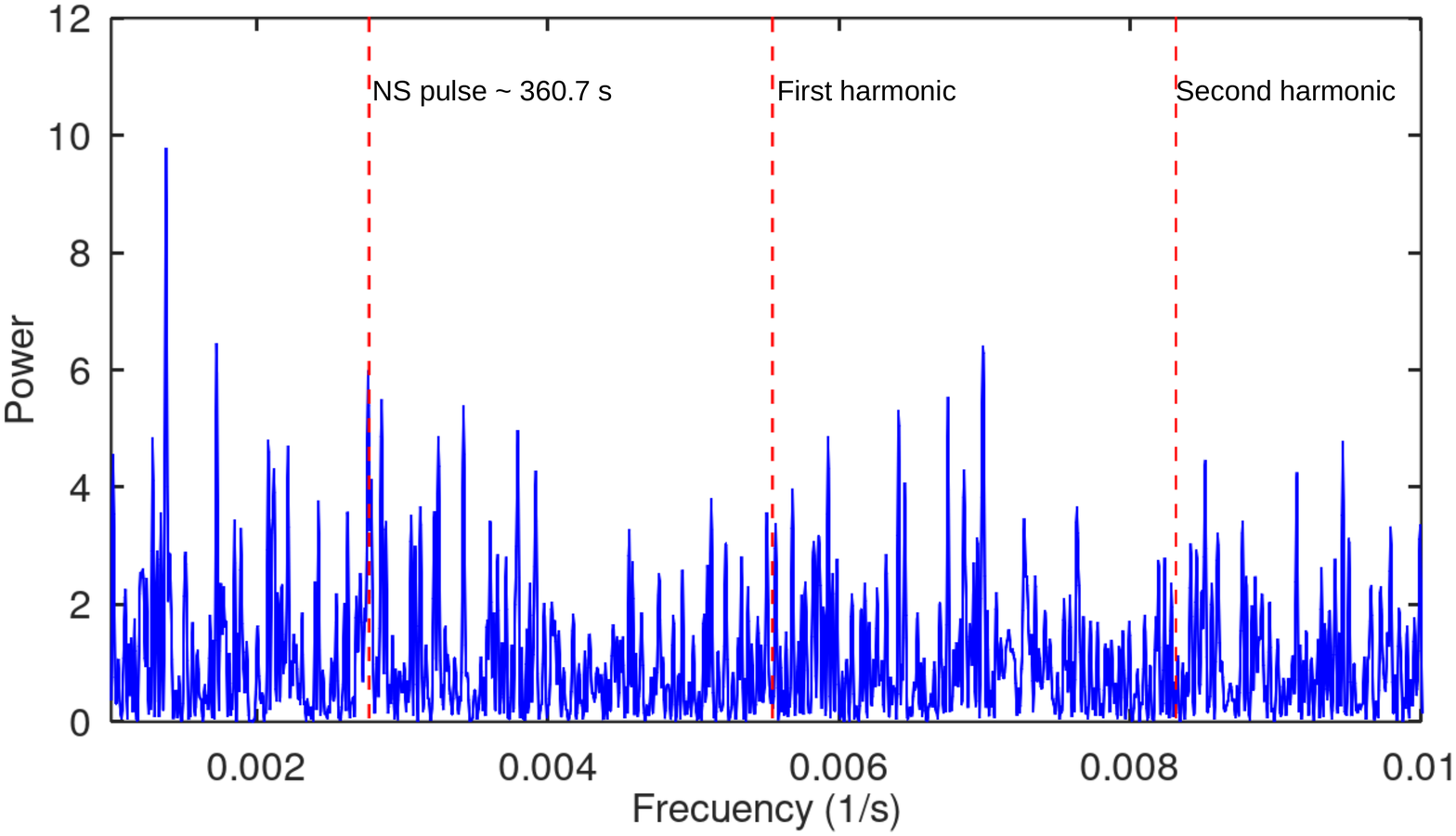}}
\caption{ \textit{Lomb Scargle} periodogram for the three EPIC cameras combined lightcurve. } 
\label{periodogram}
\end{figure}

We searched for the NS spin pulse, with negative results. In Fig. \ref{periodogram} we present the resulting Lomb-Scargle periodogram. The expected frequencies of the pulse and subsequent harmonics have been marked. No significant signal is revealed.

\begin{figure}
\subfigure{\includegraphics[trim={2cm 2cm 2cm 2cm},width=0.9\columnwidth]{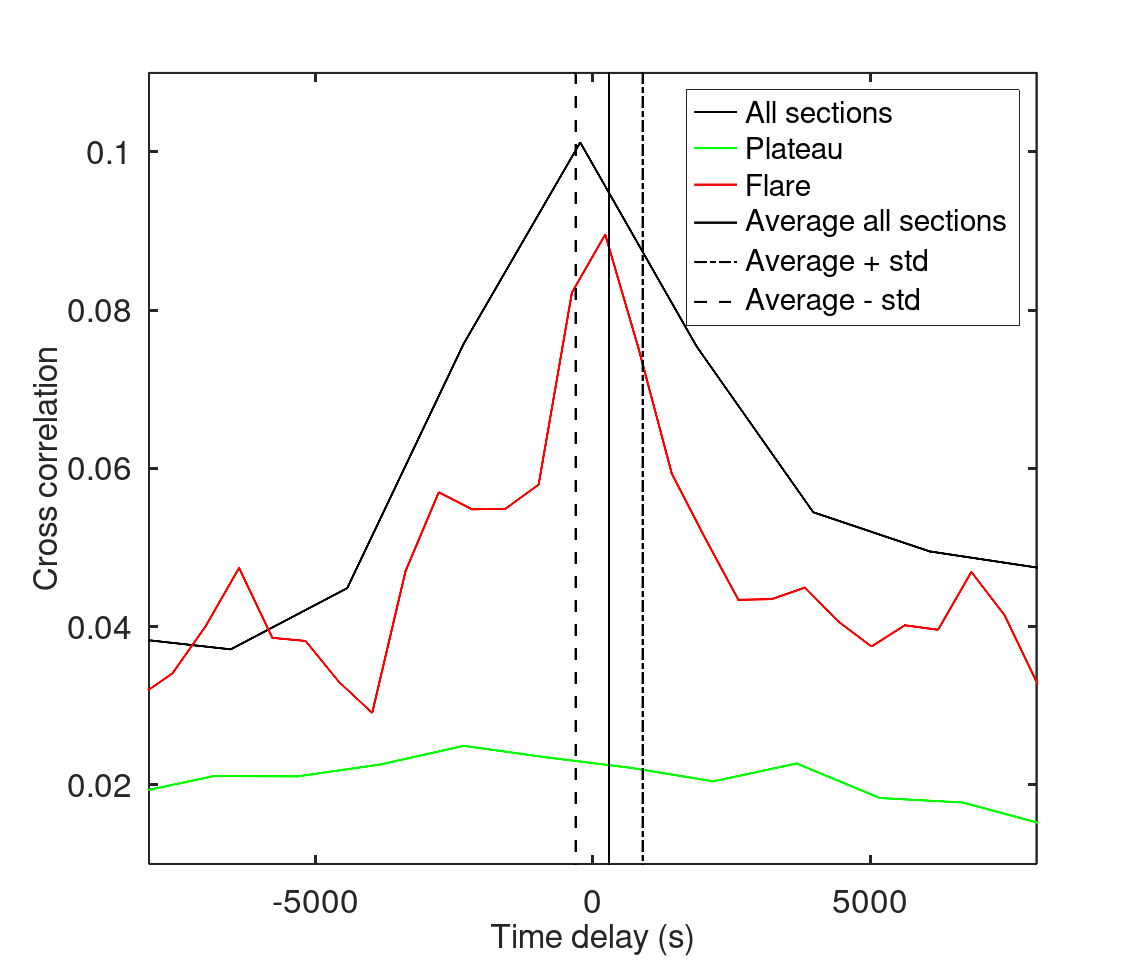}}
\caption{Example of one of the 5000 cross correlation calculations performed to derive the time delay. A 150 s time bin was used in this particular case. The average time delay obtained for the whole light curve, from the 5000 iterations, plus-minus its standard deviation, are represented with vertical black lines. } 
\label{delay}
\end{figure}

We searched for a time delay between the Fe K$\alpha$ line and the hard continuum. To this end, we produced the line ($6.25-6.55$ keV) and the hard continuum ($7-12$ keV) lightcurves, for the flare, plateau and the whole lightcurve, respectively and applied the cross-correlation method described in \cite{ding2021timing}. To obtain a reliable value of the time delay, the cross correlation was calculated 5000 times with different random lightcurve re-samplings. The result can be seen in Fig. \ref{delay}. No relevant time delay was found. The obtained time delay was 40 $\pm$ 160 s for the flare section and 300 $\pm$ 600 s for the whole lightcurve, both compatible with 0.

\subsection{Spectra: phenomenological model}
\label{res:sec:spectra_pg}

\begin{table*}													
\centering													
\caption{Phenomenological model continuum (absorbed powerlaw) spectral parameters (Fig. \ref{pow_plot}).}								
\begin{tabular}{lcccccc}													
\toprule											

&		Average	&	Flare	&	Plateau	&	High plateau	&	Low plateau	\\
\midrule
  $ \chi ^2 $ 	  	&		1.32	&	1.19	&	1.21	&	1.09	& 0.99		\\\\
  $N_{\rm H,1}$ 	 ($\times10^{22}$ cm$^{-2}$) 	& $6.9\pm0.4$	& $5.7\pm0.4$	& $7.1^{+0.6}_{-0.4}$	&  $6.9^{+0.5}_{-0.4}$	&	   $8.4^{+0.8}_{-0.6}$	\\\\
 $C$ 	  	&	 $1.00^{+0.01}_{-0.12}$	&	   $1.00^{+0.01}_{-0.10}$	&	   $0.7^{+0.3}_{-0.2}$	&	   $1.00^{+0.01}_{-0.17}$	&	   $0.8^{+0.2}_{-0.6}$	\\\\
  $N_{\rm H,2}$ 	 ($\times10^{22}$ cm$^{-2}$) 	&		   $39 \pm 11$	&	   $28^{+18}_{-14}$	&	   $40^{+30}_{-20}$	&	   $36^{+15}_{-13}$	&	   $\left(1.0^{+0.7}_{-1.0}\right)\times10^{2}$	\\\\
 $K_{\rm po}$ 	  ($\times 10^{-5}$ ph keV$^{-1}$ cm $^{-2}$s$^{-1}$ ) 	&		   $3.4\pm0.4$	&	   $4.8^{+0.9}_{-1.0}$	&	   $2.8\pm0.4$	&	   $2.8\pm0.4$	&	   $0.41^{+0.10}_{-0.13}$	\\\\
 Flux 	 ($\times 10^{-13}$ erg cm$^{-2}$ s$^{-1}$) 	&		$7.6 \pm 0.9$	&	$13_{-2}^{+3}$	&	$5.6 \pm 0.8$	&	$6.3\pm 0.9$	&	$1.6_{-0.4}^{+0.5}$	\\\\

\hline

\bottomrule													
\end{tabular}													
\label{pow_spec}													
\end{table*}

\begin{figure}
\subfigure{\includegraphics[trim={2cm 3cm 0cm 0cm},width=1\columnwidth]{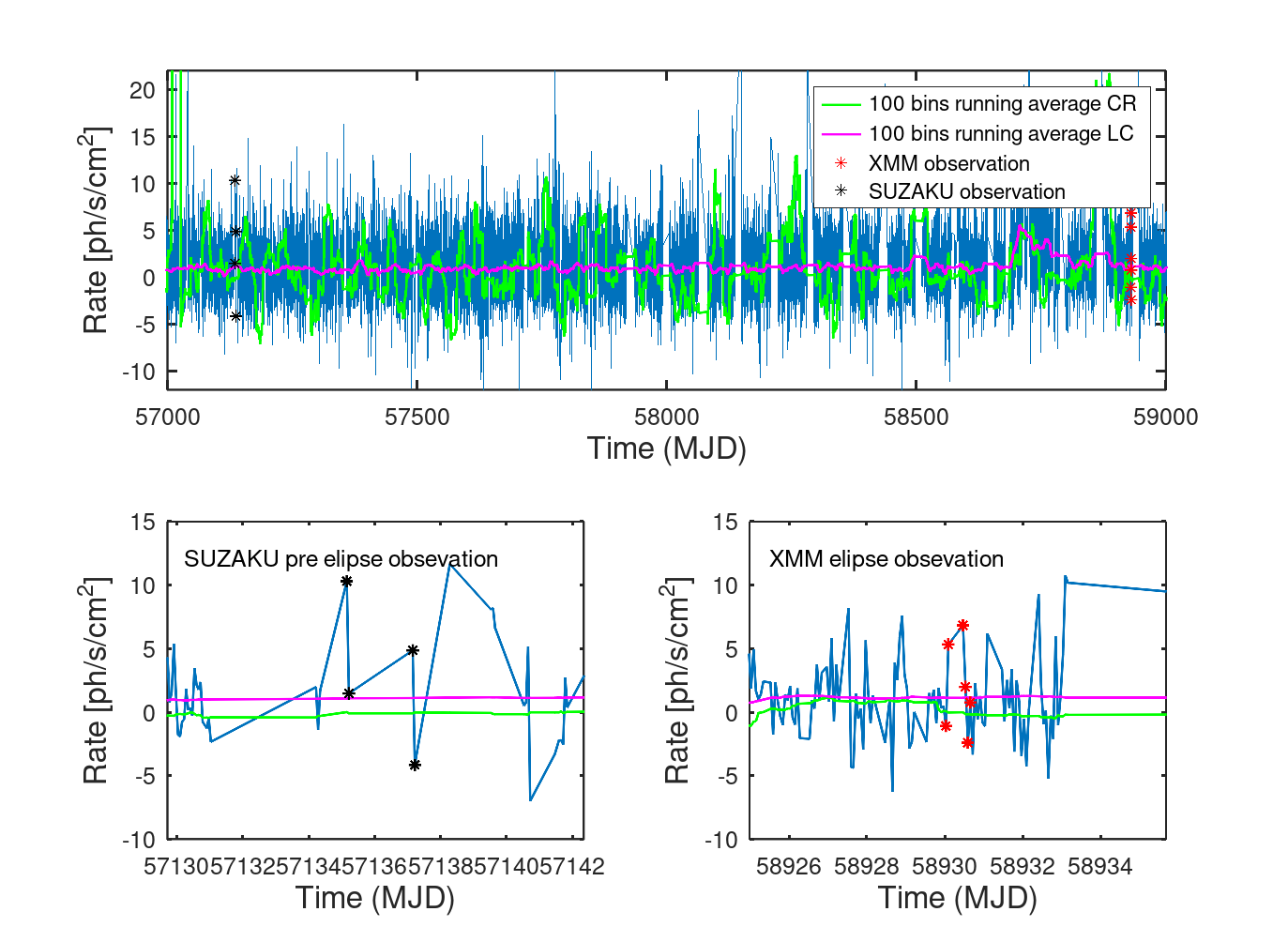}}
\caption{\textit{MAXI} {long term light curve (upper panel) containing both observations. The lower panel shows a zoom over the pre eclipse (left) and eclipse (right) observations. Black and red asterisks correspond to the \textit{Suzaku} and \textit{XMM-Newton} observations, respectively. Magenta represents the 100 bin running average light curve count rate and green represents the 100 bin running average CR.} } 
\label{maxi}
\end{figure}

We initially modeled the continuum using a black body plus a powerlaw. This is the same model as used by \citet{2018JApA...39....7D} for analysis of the {\it Suzaku} data taken just before the X-ray eclipse (pre eclipse, from now on). Besides the interstellar medium (ISM) absorption, we also allowed the presence of a local absorber, modulated by a partial covering fraction $C$ which acts as a proxy for the degree of clumping in the stellar wind of the donor star. The ISM absorption is modeled by the X-ray absorption model Tuebingen-Boulder \texttt{tbnew}. This model calculates the cross section for X-ray absorption by the ISM as the sum of the cross sections due to the gas-phase, the grain-phase, and the molecules in the ISM \citep{2000ApJ...542..914W}. For the \textit{XMM-Newton} observation analysed here, the black body component was clearly negligible. The best fit was achieved by a simple photon powerlaw where $\Gamma$ is a dimensionless photon index and the normalization constant, $K$, are the spectral photons keV$^{-1}$ cm$^{-2}$ s$^{-1}$ at 1 keV. 

The model used is described by Eq. \ref{model}.

\begin{equation}
\label{model}
\begin{split}
F(E) & = [\exp(-N_{\rm H,1}\sigma(E))+\\
& +C\exp\left(-N_{\rm H,2}\sigma(E) \right)] \left[\texttt{$KE^{-\Gamma}$ + G}\right ]
\end{split}
\end{equation} 

\begin{figure*}
\subfigure{\includegraphics[trim={5cm 2.3cm 5cm 0.67cm},width=0.75\textwidth]{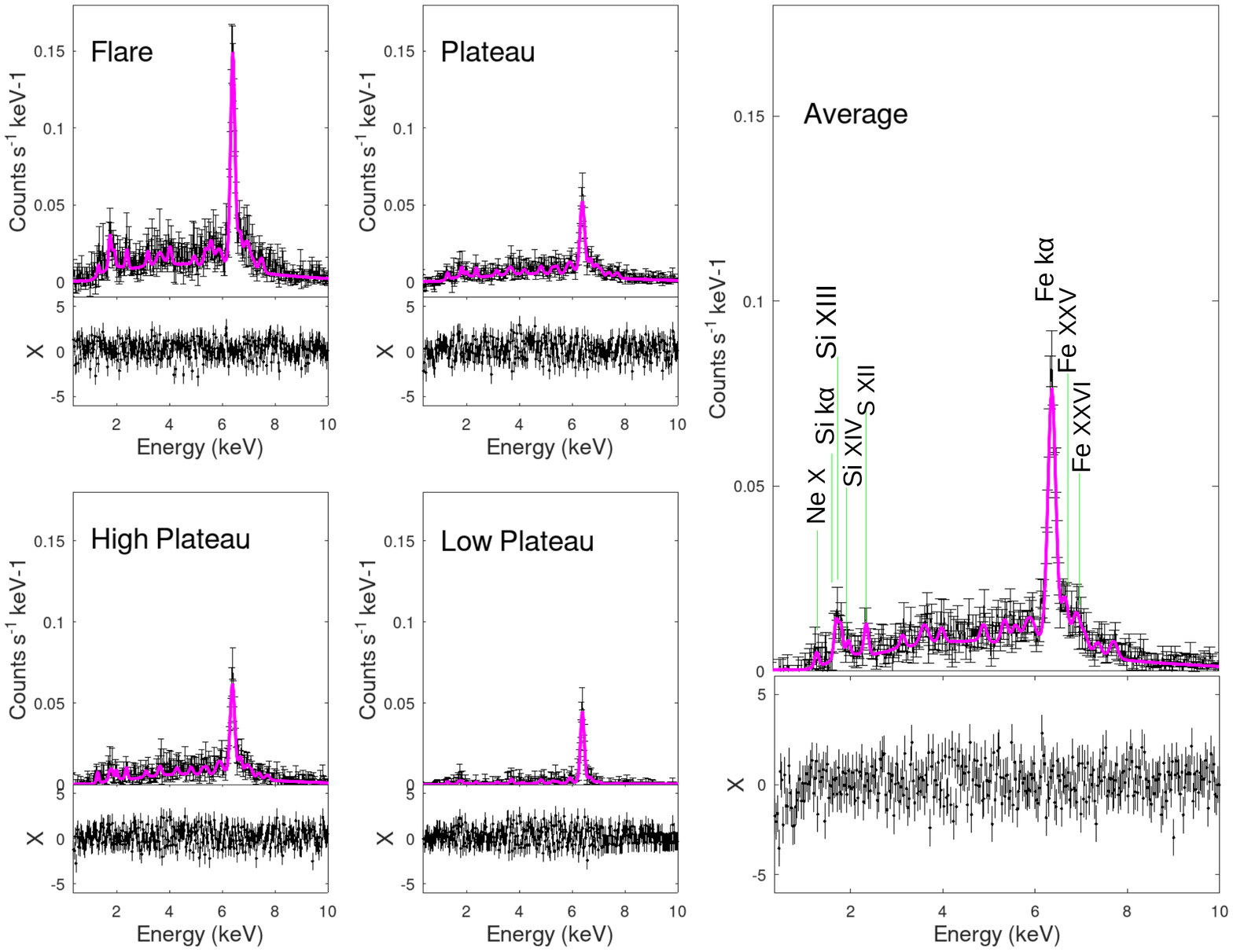}}
\caption{Phenomenological model (powerlaw plus gaussians) fit to the eclipse spectrum: data (black) and model (magenta) for the average and flux resolved spectra, respectively. All spectra are represented in the same scale to appreciate variability. Only the most relevant lines detected are marked in the average spectrum. The corresponding data are in Tables \ref{pow_spec}, continuum, and \ref{strong_lines}, lines).} 
\label{pow_plot}
\end{figure*}

Where $G$ represents the Gaussian functions added to account for the emission lines. Modelling the continuum during an eclipse is complicated because it is strongly suppressed and it is dominated by emission lines. To model the powerlaw correctly, we have used the \emph{Monitor of All-sky X-ray Image} (\textit{MAXI})\footnote{http://maxi.riken.jp} long term lightcurve of the source \citep{10.1093/pasj/61.5.999}, plotted in Fig. \ref{maxi}, with both, \emph{Suzaku} and \emph{XMM-Newton} observations, marked. To help the analysis, we over plot the 100 bin running average for both the lightcurve and the color ratio ($CR=(4-10){\rm keV}/(2-4){\rm keV}$). No clear differences seem to exist between the two epochs. The source appears to be in the same long term state. Besides the photoelectric absorption, X-ray photons are also scattered off the stellar wind. For the energies involved ($E<10$ keV) the scattering is in the Thomson regime, with no energy dependence, that is to say, conserving the continuum spectral shape. When let to vary free, the photon index turns out to be $\Gamma=0.5$, harder than during pre eclipse. However, as stated before, the general state of the source appears to be essentially the same in both observations. Therefore we fixed the powerlaw photon index to the pre eclipse value ($\Gamma=1.12$) in all our spectra. This worsened the average $\chi^2$ fit by only 1.5\%. $N_{\rm H.2}$ ranged from 36 to 45, compatible with our model. The largest difference ($\sim 17\%$) was is $N_{\rm H.1}$, which ranged between 5.46 and 5.95. In any case, the obtained plasma parameters derived below are not sensitive to the photon index value and in general remain compatible within the uncertainties. The best fit parameters are presented in Tables \ref{pow_spec} (continuum) and \ref{strong_lines} (lines). The spectra are presented in Fig. \ref{pow_plot}. 

Apart from the lack of the blackbody component, other differences arise when comparing with the pre eclipse observation \citep{2018JApA...39....7D}. Here, the covering fraction is $C\sim 1$ for all phases while it was 0.68 during pre eclipse. 

The powerlaw norm was a hundred times lower during the eclipse. For a given distance, the absorption corrected  fluxes of Table \ref{res:sec:spectra_pg} translate into an X-ray luminosity ratio of $\sim 70$ between pre and eclipse\footnote{$L_{\rm X}^{\rm eclipse}\approx 1.5\times 10^{34}$ erg s$^{-1}$ for $d$ in Table \ref{parameters}.}. Such ratio, although large, is well within the range found for eclipsing SGXBs \citep[][their Table 6]{2019ApJS..243...29A}. 

In Table \ref{strong_lines} we list the strongest lines found. The Fe lines intensities respond positively to the continuum illumination. The intensity of the Fe K$\alpha$ line, on the average spectrum, $I_{\rm FeK\alpha}=(10.4\pm 0.9) \times 10^{-6}$ ph s$^{-1}$ cm$^{-2}$, is $\sim 7.4$ times lower than the one measured during pre eclipse, $I_{\rm FeK\alpha}=77 \times 10^{-6}$ ph s$^{-1}$ cm$^{-2}$ \citep{2018JApA...39....7D}. The presence of this emission from near neutral Fe together with the highly ionized species, He-like \ion{Fe}{xxv} and H-like \ion{Fe}{xxvi}, means that the observed spectrum comes from gas in two phases, with very low and very high ionization, respectively. \ion{Fe}{xxvi}/\ion{Fe}{xxv} ratios are $\sim  0.91, 0.59, 0.70, 0.66$ and
0.33, for the average and flux resolved spectra, respectively. These values are compatible with a high ionization
parameter $\log\xi\geqslant 3.4$ \citep[][their Table 5]{1996PASJ...48..425E}. However, \ion{Fe}{xxv}/\ion{Si}{xiv} ratios, namely 0.67, 0.40, 0.64, 0.46 and 1.88, are compatible with a plasma with an ionization
parameter lower than 2.4 \citep[][]{1996PASJ...48..425E}.


\subsection{Spectra: plasma emission code \texttt{photemis}}
\label{res:sec:spectra_phot}

\begin{figure*}
\subfigure{\includegraphics[trim={5cm 1cm 5cm 0cm},width=0.75\textwidth]{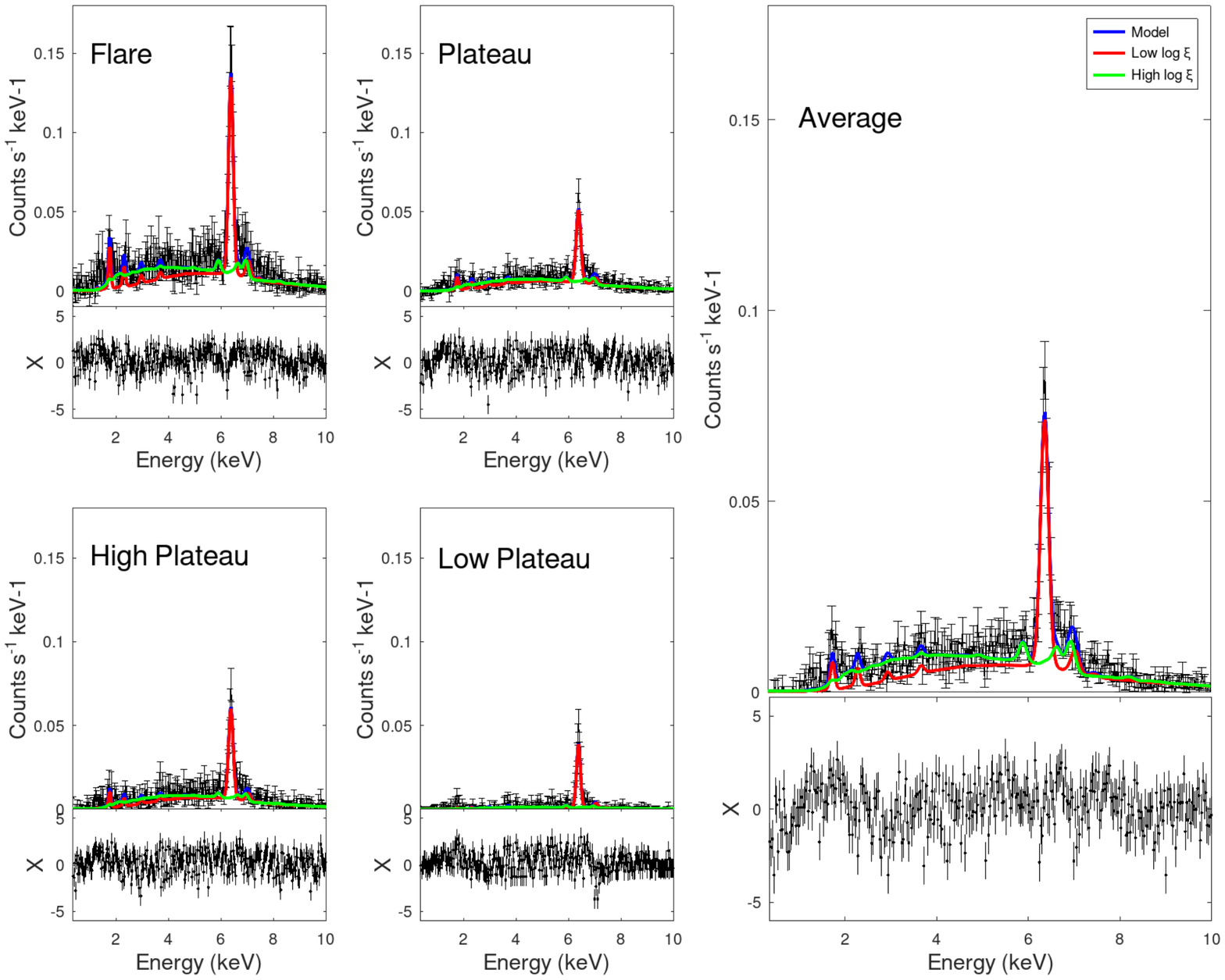}}
\caption{\texttt{photemis} model of the eclipse spectra: blue is the model, red the low ionised component and green the highly ionised component. The right panel presents the average spectrum. In the left panel, from top left to bottom right, the flare, plateau, low plateau and high plateau spectra are represented at the same scale to appreciate variations. The corresponding parameters are summarised in Table \ref{phot_spec}.} 
\label{phot_plot}
\end{figure*}

Apart of the phenomenological model described in the preceding section, we have used a self consistent plasma emission code. For that purpose we use \texttt{photemis}. This model describes the thermal (i.e. recombination and collisional excitation) emission which comes from a plasma,  using the  \textsc{xstar} code \citep{2001ApJS..133..221K}, without including the resonant scattered line emission. The model supplies the emissivity of the gas, in units of erg cm$^{-3}$ s$^{-1}$. The model used was:

\begin{equation}
\label{model2}
\begin{split}
F(E) & = [\exp(-N_{\rm H,1}\sigma(E))+\\
& +C\exp\left(-N_{\rm H,2}\sigma(E) \right)] \left[\texttt{photemis$_{1}$+photemis$_{2}$+powerlaw}\right ]
\end{split}
\end{equation} 

where a powerlaw is used to describe the continuum and \texttt{photemis} the pure emission line spectrum. Two \texttt{photemis} components, with low and high ionization parameters ($\xi$) respectively, are required to describe the main emission lines. The best fit parameters are presented in Table \ref{phot_spec} and the corresponding data plus the fitted model in Fig. \ref{phot_plot}. As in the phenomenological model (see Sec. \ref{res:sec:spectra_pg}) the photon index value was set to the reported pre eclipse value ($\Gamma=1.12$).

\begin{table*}												
\centering												
\caption{\texttt{photemis} model spectral parameters (Fig. \ref{phot_plot}).}												
\begin{adjustbox}{max width=\textwidth}												
\begin{tabular}{lcccccc}												
\toprule												
		&	Average	&	Flare	&	Plateau	&	High plateau	&	Low plateau	\\
\midrule
  $ \chi ^2 $	&1.53		&	1.23	&	1.45	&	1.3	& 1.06		\\\\
 $N_{\rm H.1}$	&	$8.0\pm 0.4$	&	$7.1^{+0.6}_{-0.3}$	&	$7.7^{+0.6}_{-0.5}$	&	$7.9^{+0.7}_{-0.6}$	&	$11^{+2}_{-2}$	\\\\
$C$	&	$1.00^{+0.01}_{-0.05}$	&	$1.00^{+0.01}_{-0.08}$	&	$1.00^{+0.01}_{-0.07}$	&	$1.00^{+0.01}_{-0.07}$	&	$1.00^{+0.01}_{-0.15}$	\\\\
 $N_{\rm H.2}$	&	$44^{+6}_{-5}$	&	$51^{+9}_{-8}$	&	$40^{+7}_{-6}$	&	$41^{+7}_{-6}$	&	$28^{+12}_{-7}$	\\\\
$K_{\rm po}$	&	$4.2 \pm 0.2$	&	$6.6\pm 0.4$	&	$3.1\pm 0.2$	&	$3.4 \pm 0.2$	&	$0.38^{+0.07}_{-0.07}$	\\\\
$K_{\rm phot_{1}}$	&	$4000 \pm 300$	&	$8000 \pm 700 $	&	$2700 \pm 240$	&	$3300\pm 300$	&	$1900^{+120}_{-240}$	\\\\
log($\xi_{1})$	&	$0.36^{+0.01}_{-0.08}$	&	$0.36^{+0.01}_{-0.08}$	&	$0.36^{+0.01}_{-0.08}$	&	$0.36^{+0.01}_{-0.08}$	&	$0.36^{+0.01}_{-0.08}$	\\\\
$K_{\rm phot_{2}}$	&	$2.9 \pm 0.5$	&	$5.8^{+1.9}_{-1.0}$	&	$2.0 \pm 0.5$	&	$2.3 \pm 0.6$	&	$0.5^{+0.2}_{-0.3}$	\\\\
log($\xi_{2})$	&	$3.30 \pm 0.08$	&	$3.40^{+0.07}_{-0.10}$	&	$3.30 \pm 0.07$	&	$3.30\pm 0.07$	&	$3.0\pm 0.2$	\\\\
$v_{\rm turb}$	&	$2400^{+600}_{-1000}$	&	$2100^{+900}_{-1800}$	&	$2800^{+200}_{-1300}$	&	$2600^{+400}_{-1400}$	&	$300^{+400}_{-10}$	\\\\
\hline	
\bottomrule												
\end{tabular}												
\end{adjustbox}												
\label{phot_spec}												
\end{table*}												

The two plasmas have low ($\log\xi_{1}\approx 0.36$) and high ($\log\xi_{2}\approx 3.7$) ionization states. The fit requires broadening of the lines with a corresponding turbulence velocity of $v_{\rm turb}\approx 3000$ km s$^{-1}$. This velocity was tied during the fits to have the same value for both plasmas. The normalization of the cold plasma ($K_{\rm phot_{1}}$) is much larger than the one for the hot plasma ($K_{\rm phot_{2}}$). The physical meaning of the normalization  ($K_{\rm phot}$), is:

\begin{displaymath}
K=\frac{{EM}}{4\pi d^2} \times 10^{-10}
\end{displaymath}


where $EM=\int{n_{\rm e}n_{\rm i}dV\approx n^{2}V}$ is the emission measure of the gas (at the ionization parameter used in the fit) and $d$ is the distance to the source. Therefore the two gas phases have an emission measure ratio $EM_{1}/EM{_2}=EM_{\rm cold}/EM_{\rm hot}\approx 10^{3}$. For a source distance of $d\approx 7.4$ kpc, the $EM_{\rm cold}\approx 3\times 10^{59}$ cm$^{-3}$ and $EM_{\rm hot}\approx 2\times 10^{56}$ cm$^{-3}$. These values are in agreement with those found for other SGXRBs \citep[i.e.][]{2021MNRAS.501.5646M}. 
\section{Discussion}

The comparison of the eclipse and pre eclipse spectra allows to extract interesting conclusions. During eclipse, the observed spectrum is the sum of the scattered radiation plus the intrinsic X-ray emission from the donor star. OB stars have X-ray luminosities of the order of $10^{32}$ erg s$^{-1}$ \citep[v.g.][]{2018A&A...620A..89N}, 100 times lower than observed here and, also, soft thermal spectra with $kT\sim 0.1-0.2$ keV. Therefore, the observed EPIC spectrum ($0.35-10$ keV) is clearly dominated by the scattered component. This is consistent with the time delay found, compatible with zero, as both observed components, Fe K$\alpha$ line and hard continuum, are reflected (scattered) in the donor's wind, during eclipse. 

The blackbody component of the pre eclipse model, with a temperature of $kT_{\rm bb}=0.12$ keV, used to describe the soft excess at low energies, is not detected during eclipse. This rules out its origin as the stellar wind of the donor. It has to be produced close to the NS or at the accretion stream along the line connecting the NS and the donor. 

\begin{figure}
\subfigure{\includegraphics[trim={2cm 0cm 2cm 0cm},width=0.9\columnwidth]{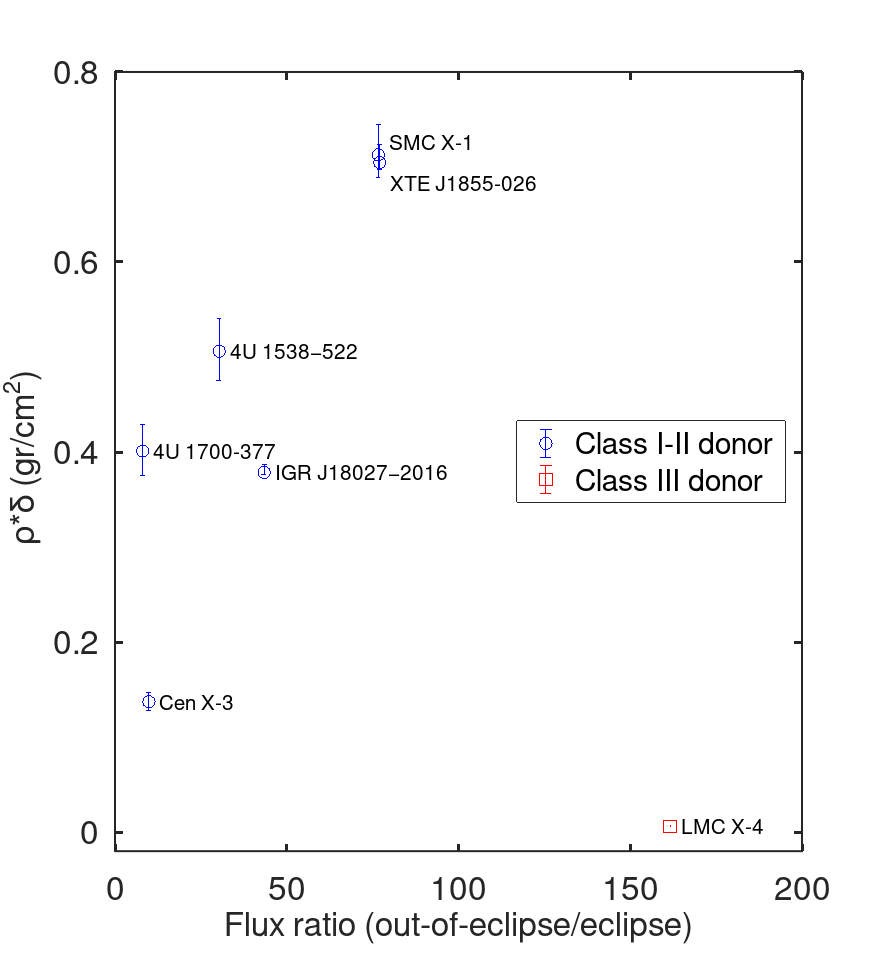}}
\caption{$\delta$ represents the difference in path traveled by a photon emitted at orbital phase 0 and a photon emitted in the closest-to-observer phase. This magnitude, multiplied by the wind density along this path, $\rho(r)$, is depicted vs flux ratio between eclipse and out-of-eclipse observations within the [$0.3-10.0$] keV energy range.} 
\label{fig:ratio}
\end{figure}

The powerlaw photon index during eclipse is compatible with the pre eclipse value. However the  absorption corrected $L_{\rm X}^{\rm eclipse} \approx 1.5\times 10^{34}$ erg s$^{-1}$  is 70 times lower than pre eclipse ($L_{\rm X}\approx 1.0\times 10^{36}$ erg s$^{-1}$). Such ratio is rather large albeit well within those found in eclipsing SGXBs \citep[][their Table 6]{2019ApJS..243...29A}. To further explore this issue, we have compiled data from some eclipsing systems in Table \ref{ratio}. On this Table, $\delta$ refers to the difference in path traveled by an X-ray photon emitted at orbital phase 0 and a photon emitted in the closest-to-observer phase. In order to calculate this distance the semi major axis, the eccentricity, inclination and argument of the periapsis, were taken into account. In Fig.\ref{fig:ratio} we plot the wind density integrated along this path versus the flux ratio (column density). Two conclusions can be extracted. First, there is a positive trend for all systems with class I-II donors (LMC X-4 is class III), indicating that absorption is a major driver of the observed flux ratio. Second, our source XTE J1855$-$026 is high on this trend but within the normal values displayed by eclipsing HMXBs. Deep X-ray eclipses, which allow for large out-of-eclipse to eclipse luminosity ratios, are possible provided that the wind occulted by the donor star (the X-ray shadow) is unionised, so that every scattered photon entering it, is locally absorbed \citep[][their Fig. 4, corresponding to model 2a]{1978ApJ...224..614H}.

\begin{figure}
\subfigure{\includegraphics[trim={0cm 5cm 0cm 0cm},width=0.9\columnwidth]{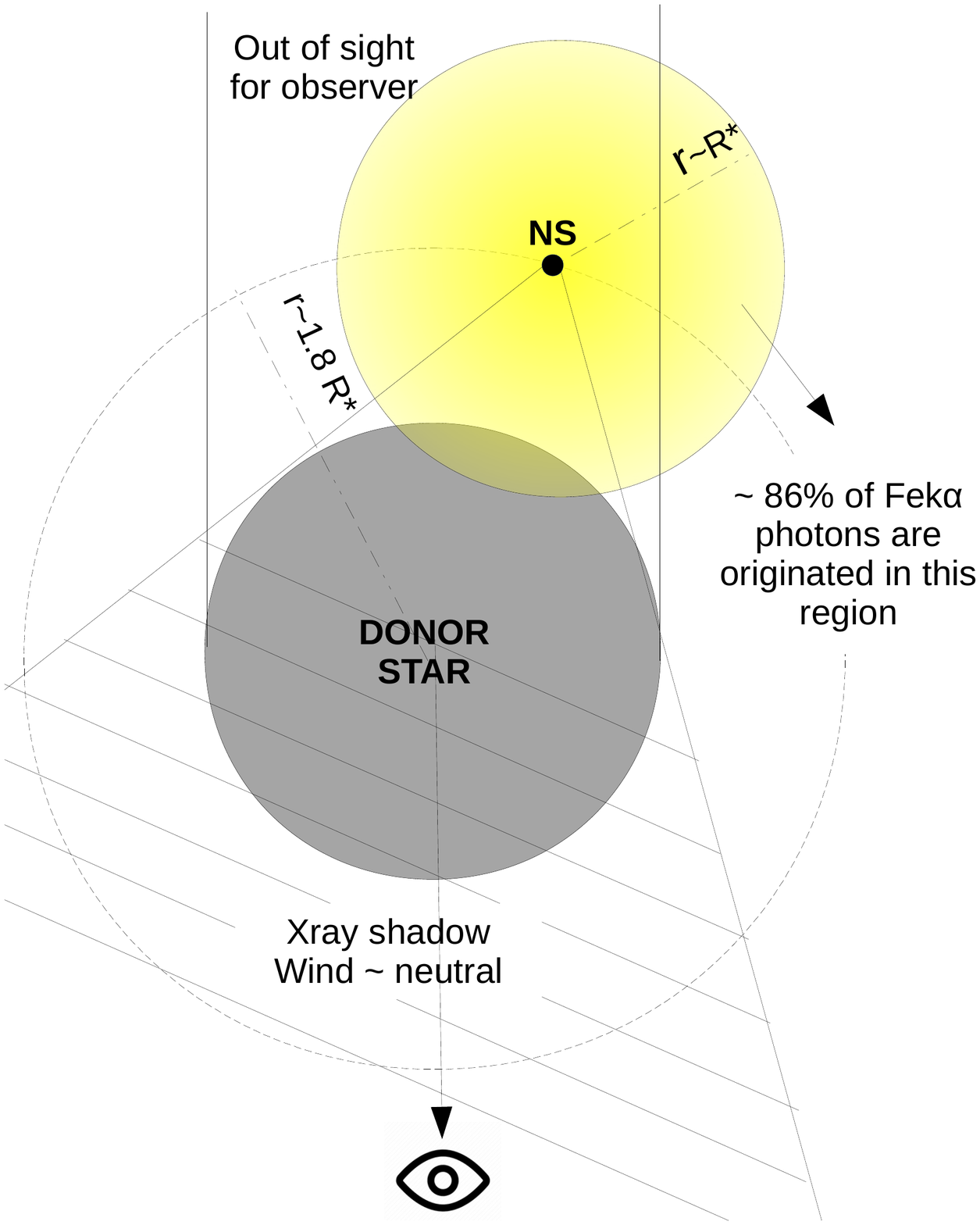}}
\caption{Scheme of the system. The orbit, donor star and the Fe K$\alpha$ emitting region are to scale. } 
\label{scheme}
\end{figure}

The intensity of the Fe K$\alpha$ line, on the average spectrum, $I_{\rm FeK\alpha}=(10.4\pm 0.9) \times 10^{-6}$ ph s$^{-1}$ cm$^{-2}$, is $\sim 7.4$ times lower than the one measured during pre eclipse, $I_{\rm FeK\alpha}=77 \times 10^{-6}$ ph s$^{-1}$ cm$^{-2}$ \citep{2018JApA...39....7D}. Fe K$\alpha$ photons can not be resonantly scattered in the wind because they do not have the required energy ($E_{\rm K edge}>7.112$ keV) to induce further fluorescence. Therefore, these photons must be produced in the direct line of sight towards the observer and the NS, simultaneously. This means that the vast majority of Fe K$\alpha$ emission must come from distances $r_{\rm X}<1R_{*}$ from the NS (Fig. \ref{scheme}). 

As explained in section \ref{res:sec:spectra_pg}, two plasmas, at different ionization states, are required to describe the eclipse spectrum. These two gas phases can be identified with the clumped part of the wind (cold and dense) and the inter clump medium (hot and rarefied). The observed ratio $EM_{1}/EM{_2}=EM_{\rm cold}/EM_{\rm hot}\approx 10^{3}$ allows us to compute the density contrast $n_{\rm c}/n_{\rm i}$ between the clump and interclump gas phases. Indeed, 

\begin{equation}
    EM_{\rm cold}/EM_{\rm hot}\approx\left(\frac{n_{\rm c}}{n_{\rm i}}\right)^{2}\left(\frac{V_{\rm c}}{V_{\rm i}}\right)
\end{equation}

where $V_{\rm c}$ and $V_{\rm i}$ are the wind volumes occupied by the clumps and the interclump medium, respectively, and $V_{\rm wind}=V_{\rm c}+V_{\rm i}$. This can be expressed as a function of the clumps volume filling factor, $f_{\rm V}=V_{\rm cl}/V_{\rm wind}$, as

\begin{equation}
    EM_{\rm cold}/EM_{\rm hot}=\left(\frac{n_{\rm c}}{n_{\rm i}}\right)^{2}\frac{f_{\rm V}}{1-f_{\rm V}}
\end{equation}

Now, assuming $f_{\rm V}\approx [0.04-0.05]$ \citep[][for the cases of Vela X-1, B0.5I, $f_{\rm V}\approx 0.04$ and QV Nor, O6.5I, $f_{\rm V}\approx 0.05$, respectively] {1999ApJ...525..921S, 2021MNRAS.501.5646M}, we get $n_{\rm c}/n_{\rm i}\approx 180$, in line with expectations from stellar wind models for massive stars \citep{Oskinova_2011, Hainich_2020}.

\section{Summary and conclusions}

We present the first X-ray observation of the HMXB XTE J1855$-$026 taken entirely during the eclipse of the NS. This allows us to a) compare with the parameters obtained during the existing pre eclipse observation and b) explore the back illuminated stellar wind of the B0I donor. The main conclusions are: 

\begin{itemize}
    \item The black body component, used to describe the soft excess during pre eclipse, is not observed during eclipse. It must be then produced near the NS or along the donor-NS line. 
    
    \item The $0.3-10$ keV luminosity during eclipse ($\sim 10^{34}$ erg s$^{-1}$) is 70 times lower than pre eclipse, well within the range found for eclipsing SGXRB's. This large ratio would not be due to a different state of the source, as suggested by the long term light curve, but to deeper X-ray eclipses caused by the absorption of scattered photons in the non illuminated part of the wind. 
    
    \item The intensity of the Fe K$\alpha$ line, on the average eclipse spectrum, is $\sim 7.4$ times lower than the one measured during pre eclipse. Since K$\alpha$ photons can not be resonantly scattered in the wind, the vast majority of Fe K$\alpha$ emission must come from distances $r_{\rm X}<1R_{*}$ from the NS. 
    
    \item The eclipse spectrum is successfully modelled through the addition of two photoionized plasmas, one with low ionization ($\log\xi_{\rm 1,cold}=0.36$) and high emission measure ($EM_{\rm 1,cold}\approx 3\times 10^{59}$ cm$^{-3}$) and another with high ionization ($\log\xi_{\rm 2,hot}=3.7$) and low emission measure ($EM_{\rm 2,hot}\approx 2\times 10^{56}$ cm$^{-3}$).
    
    \item Assuming that the cold and hot gas phases are the clumps and the interclump medium of the stellar wind, respectively, and a clump volume filling factor $f_{\rm V}\approx [0.04-0.05]$, as observed for massive stars, a density contrast between clumps and the interclump medium of $n_{\rm c}/n_{\rm i}\approx 180$ is deduced in agreement with theoretical expectations and optical-UV observations of massive star winds. 
    
\end{itemize}

\section*{Acknowledgements}
We thank the referee for the constructive criticism and valuable suggestions and Dr. Lida Oskinova for important discussions on massive star stellar winds. This research has made use of a collection of ISIS functions (ISISscripts) provided by ECAP/Remeis observatory and MIT (\url{http://www.sternwarte.uni-erlangen.de/isis/}). This work has made use of data from the European Space Agency (ESA) mission {\it Gaia} (\url{https://www.cosmos.esa.int/gaia}), processed by the {\it Gaia} Data Processing and Analysis Consortium (DPAC, \url{https://www.cosmos.esa.int/web/gaia/dpac/consortium}). Funding for the DPAC has been provided by national institutions, in particular the institutions participating in the {\it Gaia} Multilateral Agreement.

\section*{Data Availability}

The observation can be found in the \textit{XMM-Newton} archive under the observation ID 0844630101.



\bibliographystyle{mnras}
\bibliography{example} 



\appendix

\section{Strong emission lines}

\begin{table*}																	
\centering																	
\caption{Parameters of the strongest lines found in the average and flux resolved spectra (Fig. \ref{pow_plot}) (\textsc{atomdb} data base).}																
\begin{adjustbox}{max width=\textwidth}																	
\begin{tabular}{llccccccc}																	
\toprule																	

	&		&	Average	&	Flare	&	Plateau	&	High plateau	&	Low plateau	\\
\midrule
\ion{Ne}{x}	&	$I$ ($\times 10^{-6}$  ph s$^{-1}$ cm$^{-2}$)	&	$110_{-80}^{+120}$	&	$90_{-50}^{+60}$	&	$370_{-200}^{+300}$	&	$340_{-200}^{+300}$	&	$270_{-190}^{+240}$	\\\\
	&	Center (keV)	&	$1.32\pm 0.07$	&	$1.40_{-0.03}^{+0.04}$	&	$1.30_{-0.04}^{+0.03}$	&	$1.30_{-0.04}^{+0.05}$	&	$1.30_{-0.06}^{+0.02}$	\\\\
	&	Sigma (eV)	&	$20_{-20}^{+1}$	&	$20_{-20}^{+1}$	&	$20_{-20}^{+1}$	&	$20_{-20}^{+1}$	&	$1_{-0}^{+20}$	\\\\
	&	Eqw (eV)	&	$4000_{-2000}^{+4000}$	&	$2600_{-1500}^{+1600}$	&	$6000_{-3000}^{+4000}$	&	$14000_{-8000}^{+12000}$	&	$7000_{-5000}^{+6000}$	\\\\
Si K$\alpha$	&	$I$ ($\times 10^{-6}$  ph s$^{-1}$ cm$^{-2}$)	&	$10_{-3}^{+0}$	&	$10_{-4}^{+0}$	&	$10_{-4}^{+0}$	&	$10_{-4}^{+0}$	&	$8_{-4}^{+2}$	\\\\
	&	Center (keV)	&	$1.72_{-0.01}^{+0.02}$	&	$1.72\pm 0.01$	&	$1.72\pm 0.01$	&	$1.72\pm 0.01$	&	$1.72\pm 0.01$	\\\\
	&	Sigma (eV)	&	$1_{-0}^{+20}$	&	$2_{-0}^{+20}$	&	$1_{-1}^{+20}$	&	$1_{-1}^{+20}$	&	$1_{-1}^{+20}$	\\\\
	&	Eqw (eV)	&	$550_{-160}^{+0}$	&	$380_{-160}^{+0}$	&	$660_{-240}^{+0}$	&	$700_{-300}^{+0}$	&	$3900_{-2000}^{+700}$	\\\\
\ion{Si}{xiii} / \ion{Fe}{xxiv}	&	$I$ ($\times 10^{-6}$  ph s$^{-1}$ cm$^{-2}$)	&	$7 \pm 3$	&	$8_{-6}^{+2}$	&	$6_{-3}^{+3}$	&	$5_{-3}^{+4}$	&	$2_{-2}^{+2}$	\\\\
	&	Center (keV)	&	$1.83_{-0.02}^{+0.01}$	&	$1.81_{-0.01}^{+0.02}$	&	$1.83_{-0.02}^{+0.01}$	&	$1.82_{-0.03}^{+0.01}$	&	$1.83_{-0.02}^{+0.01}$	\\\\
	&	Sigma (eV)	&	$4_{-3}^{+20}$	&	$20_{-20}^{+1}$	&	$2_{-1}^{+20.0}$	&	$1_{-0}^{+20}$	&	$2_{-1}^{+20}$	\\\\
	&	Eqw (eV)	&	$420 \pm 170$	&	$320_{-240}^{+80}$	&	$440_{-210}^{+220}$	&	$370_{-240}^{+280}$	&	$1100_{-1000}^{+1200}$	\\\\
\ion{Si}{xiii}	&	$I$ ($\times 10^{-6}$  ph s$^{-1}$ cm$^{-2}$)	&	$3_{-2}^{+2}$	&	$7_{-4}^{+4}$	&	$3 \pm 2$	&	$4_{-2}^{+2}$	&	$0.4_{-0.4}^{+0.9}$	\\\\
	&	Center (keV)	&	$1.99_{-0.05}^{+0.04}$	&	$1.93_{-0.03}^{+0.04}$	&	$2.00_{-0.04}^{+0.04}$	&	$2,00_{-0.04}^{+0.04}$	&	$2.05_{-0.15}^{+0.06}$	\\\\
	&	Sigma (eV)	&	$1_{-0}^{+20}$	&	$10_{-9}^{+10}$	&	$1_{-0}^{+20}$	&	$1_{-0}^{+20}$	&	$1_{-0}^{+20}$	\\\\
	&	Eqw (eV)	&	$180_{-110}^{+130}$	&	$310\pm10$	&	$200 \pm 130$	&	$280_{-160}^{+170}$	&	$200_{-200}^{+400}$	\\\\
 \ion{S}{xii} 	&	$I$ ($\times 10^{-6}$  ph s$^{-1}$ cm$^{-2}$)	&	$4_{-1}^{+1}$	&	$5_{-2}^{+3}$	&	$3 \pm 1$	&	$3 \pm 2$	&	$0.8\pm 0.7$	\\\\
	&	Center (keV)	&	$2.38_{-0.02}^{+0.02}$	&	$2.39_{-0.04}^{+0.02}$	&	$2.37_{-0.03}^{+0.03}$	&	$2.38_{-0.03}^{+0.03}$	&	$2.37_{-0.08}^{+0.04}$	\\\\
	&	Sigma (eV)	&	$1_{-0}^{+20}$	&	$20_{-20}^{+1}$	&	$1_{-0}^{+20}$	&	$1_{-0}^{+20}$	&	$1_{-0}^{+20}$	\\\\
	&	Eqw (eV)	&	$290_{-100}^{+110}$	&	$290_{-130}^{+140}$	&	$260 \pm 130$	&	$280 \pm 140$	&	$500 \pm 500$	\\\\
Fe K$\alpha$	&	$I$ ($\times 10^{-6}$  ph s$^{-1}$ cm$^{-2}$)	&	$10_{-1}^{+1}$	&	$20_{-6}^{+2}$	&	$8 \pm 1$	&	$9 \pm 1$	&	$8 \pm 1$	\\\\
	&	Center (keV)	&	$6.41 \pm 0.01$	&	$6.41_{-0.02}^{+0.01}$	&	$6.40 \pm 0.01$	&	$6.40\pm 0.01$	&	$6.40 \pm 0.01$	\\\\
	&	Sigma (eV)	&	$40_{-18}^{+14}$	&	$40_{-40}^{+20}$	&	$50_{-21}^{+17}$	&	$50_{-24}^{+20}$	&	$1_{-0}^{+30}$	\\\\
	&	Eqw (eV)	&	$2500_{-200}^{+180}$	&	$3200_{-900}^{+300}$	&	$2500 \pm 240$	&	$2500 \pm 300 $	&	$15300_{-1400}^{+1500}$	\\\\
\ion{Fe}{xxv}	&	$I$ ($\times 10^{-6}$  ph s$^{-1}$ cm$^{-2}$)	&	$2_{-1}^{+1}$	&	$3_{-2}^{+2}$	&	$2_{-1}^{+1}$	&	$2 \pm 1$	&	$1.0_{-0.4}^{+0.3}$	\\\\
	&	Center (keV)	&	$6.69_{-0.04}^{+0.03}$	&	$6.67_{-0.08}^{+0.05}$	&	$6.70\pm 0.04$	&	$6.70_{-0.05}^{+0.04}$	&	$6.70 \pm 0.04$	\\\\
	&	Sigma (eV)	&	$1_{-1}^{+100}$	&	$1_{-0}^{+100}$	&	$1_{-0}^{+100}$	&	$1_{-0}^{+100}$	&	$1.0_{-0.1}^{+70.0}$	\\\\
	&	Eqw (eV)	&	$500_{-200}^{+300}$	&	$500 \pm 300$	&	$470_{-220}^{+210}$	&	$500\pm 240 $	&	$1600_{-800}^{+600}$	\\\\
\ion{Fe}{xxvi}	&	$I$ ($\times 10^{-6}$  ph s$^{-1}$ cm$^{-2}$)	&	$2_{-1}^{+1}$	&	$2_{-2}^{+2}$	&	$1_{-1}^{+1}$	&	$1_{-1}^{+1}$	&	$0.3_{-0.3}^{+0.4}$	\\\\
	&	Center (keV)	&	$6.93 \pm 0.06$	&	$6.85_{-0.01}^{+0.15}$	&	$6.93 \pm 0.07$	&	$6.95_{-0.10}^{+0.05}$	&	$6.96_{-0.11}^{+0.04}$	\\\\
	&	Sigma (eV)	&	$30_{-30}^{+70}$	&	$2_{-1}^{+100}$	&	$1_{-0}^{+100}$	&	$1_{-0}^{+100}$	&	$1_{-0}^{+100}$	\\\\
	&	Eqw (eV)	&	$400_{-200}^{+300}$	&	$300\pm 300 $	&	$400_{-200}^{+50}$	&	$350_{-240}^{+180}$	&	$700_{-700}^{+1000}$\\

\bottomrule																
\end{tabular}															
\end{adjustbox}															
\label{strong_lines}
\end{table*}																	

\section{Emission lines}

Best fit parameters for all the gaussians included in the phenomenological model (see Section \ref{res:sec:spectra_pg}).

\begin{table*}									
\centering									
\caption{Tentative identification and parameters of weaker lines found in the average and flux resolved spectra.}									
\begin{adjustbox}{max width=\textwidth}									
\begin{tabular}{llccccc}									
\toprule							

&	&		Average	&	Flare	&	Plateau	&	High plateau	&	Low plateau	\\
	\midrule
\ion{Ar}{xvii}	&	$I$ ($\times 10^{-6}$  ph s$^{-1}$ cm$^{-2}$)	&	$0.5 \pm 0.5$	&	$1 \pm 1$	&	$0.3_{-0.3}^{+0.5}$	&	$0.4_{-0.4}^{+0.6}$	&	$0.1_{-0.1}^{+0.3}$	\\\\
&	Center (keV)	&	$3.17_{-0.07}^{+0.23}$	&	$3.19_{-0.07}^{+0.21}$	&	$3.15_{-0.05}^{+0.24}$	&	$3.13_{-0.03}^{+0.24}$	&	$3.17_{-0.07}^{+0.23}$	\\\\
&	Sigma (eV)	&	$1_{-0}^{+20}$	&	$1_{-0}^{+20}$	&	$1_{-0}^{+20}$	&	$1_{-0}^{+20}$	&	$1_{-0}^{+20}$	\\\\
&	Eqw (eV)	&	$50\pm 50$	&	$90 \pm 80$	&	$30_{-30}^{+60}$	&	$50_{-50}^{+80}$	&	$120_{-120}^{+240}$	\\\\

Ca K$\alpha$&	$I$ ($\times 10^{-6}$  ph s$^{-1}$ cm$^{-2}$)	&	$0.8_{-0.5}^{+0.6}$	&	$0.9_{-0.8}^{+1.0}$	&	$0.6_{-0.4}^{+0.6}$	&	$0.7_{-0.5}^{+0.5}$	&	$0.4 \pm 0.3 $	\\\\
 &	Center (keV)	&	$3.64_{-0.06}^{+0.05}$	&	$3.63_{-0.13}^{+0.17}$	&	$3.67_{-0.09}^{+0.06}$	&	$3.66_{-0.07}^{+0.05}$	&	$3.71_{-0.04}^{+0.04}$	\\\\
&	Sigma (eV)	&	$30_{-30}^{+70}$	&	$1_{-0}^{+100}$	&	$1_{-0}^{+100}$	&	$1_{-0}^{+100}$	&	$1.0_{-0.1}^{+80.0}$	\\\\
&	Eqw (eV)	&	$110_{-60}^{+70}$	&	$80_{-70}^{+120}$	&	$100_{-70}^{+90}$	&	$110\pm 80$	&	$500\pm 300$	\\\\

\ion{Ca}{xix} &	$I$ ($\times 10^{-6}$  ph s$^{-1}$ cm$^{-2}$)	&	$0.4_{-0.3}^{+0.4}$	&	$0.9_{-0.7}^{+0.8}$	&	$0.1_{-0.1}^{+0.3}$	&	$0.4 \pm 0.4$	&	$0.01_{-0.01}^{+0.20}$	\\\\
 &	Center (keV)	&	$4.01_{-0.12}^{+0.09}$	&	$4.02_{-0.12}^{+0.06}$	&	$4.2_{-0.3}^{+0.1}$	&	$4.3_{-0.4}^{+0.1}$	&	$4.0_{-0.1}^{+0.3}$	\\\\
&	Sigma (eV)	&	$1_{-0}^{+20}$	&	$10_{-9}^{+10}$	&	$1_{-0}^{+20}$	&	$20_{-20}^{+0}$	&	$16_{-15}^{+4}$	\\\\
&	Eqw (eV)	&	$50 \pm 50$	&	$90_{-70}^{+80}$	&	$20_{-20}^{+40}$	&	$3\pm 3$	&	$3\pm 3$	\\\\


Cr K$\alpha$/ \ion{Ca}{xx}&	$I$ ($\times 10^{-6}$  ph s$^{-1}$ cm$^{-2}$)	&	$0.7 \pm 0.3$	&	$0.8_{-0.8}^{+1.3}$	&	$0.4 \pm 0.4$	&	$0.6 \pm 0.4$	&	$0.2 \pm 0.2$	\\\\
 &	Center (keV)	&	$5.4_{-0.1}^{+0.1}$	&	$5.4_{-0.2}^{+0.4}$	&	$5.3_{-0.3}^{+0.5}$	&	$5.38_{-0.12}^{+0.06}$	&	$5.3_{-0.3}^{+0.5}$	\\\\
&	Sigma (eV)	&	$20_{-20}^{+1}$	&	$1_{-0}^{+20}$	&	$1_{-0}^{+20}$	&	$20_{-20}^{+1}$	&	$1_{-0}^{+20}$	\\\\
&	Eqw (eV)	&	$130 \pm 60$	&	$120_{-110}^{+180}$	&	$140_{-140}^{+150}$	&	$140\pm 90$	&	$400\pm 400$	\\\\

Cr K$\beta$/ Mn K$\alpha$ &	$I$ ($\times 10^{-6}$  ph s$^{-1}$ cm$^{-2}$)	&	$1.1_{-0.3}^{+0.8}$	&	$1.9_{-1.2}^{+0.9}$	&	$1.5 \pm 0.5$	&	$1.1_{-0.4}^{+0.8}$	&	$0.6 \pm 0.2$	\\\\
&	Center (keV)	&	$5.94_{-0.08}^{+0.05}$	&	$5.88_{-0.12}^{+0.12}$	&	$5.93_{-0.04}^{+0.04}$	&	$5.95_{-0.06}^{+0.05}$	&	$5.93_{-0.03}^{+0.03}$	\\\\
&	Sigma (eV)	&	$2_{-1}^{+100}$	&	$100_{-100}^{+1}$	&	$80_{-70}^{+30}$	&	$1_{-0}^{+100}$	&	$1_{-0}^{+70}$	\\\\
&	Eqw (eV)	&	$310_{-90}^{+220}$	&	$290_{-190}^{+130}$	&	$370_{-120}^{+120}$	&	$400_{-100}^{+300}$	&	$1100\pm 400$	\\\\

\ion{Fe}{xxvi} &	$I$ ($\times 10^{-6}$  ph s$^{-1}$ cm$^{-2}$)	&	$0.9_{-0.9}^{+0.6}$	&	$3_{-2}^{+1}$	&	$0.8_{-0.8}^{+0.6}$	&	$0.8_{-0.8}^{+0.5}$	&	$0.1_{-0.1}^{+0.4}$	\\\\
&	Center (keV)	&	$7.12_{-0.07}^{+0.08}$	&	$7.00_{-0.01}^{+0.19}$	&	$7.10_{-0.08}^{+0.09}$	&	$7.11_{-0.08}^{+0.09}$	&	$7.06_{-0.06}^{+0.14}$	\\\\
&	Sigma (eV)	&	$1_{-0}^{+20}$	&	$1_{-0}^{+20}$	&	$1_{-0}^{+20}$	&	$1_{-0}^{+20}$	&	$1_{-0}^{+20}$	\\\\
&	Eqw (eV)	&	$230_{-230}^{+150}$	&	$400_{-300}^{+200}$	&	$230_{-230}^{+170}$	&	$240_{-240}^{+160}$	&	$100_{-100}^{+400}$	\\\\

Ni K$\alpha$&	$I$ ($\times 10^{-6}$  ph s$^{-1}$ cm$^{-2}$)	&	$0.9\pm 0.4$	&	$1.4_{-1.2}^{+0.9}$	&	$0.9 \pm 0.4$	&	$0.9 \pm 0.5$	&	$0.03_{-0.03}^{+0.20}$	\\\\
	 &	Center (keV)	&	$7.4_{-0.1}^{+0.1}$	&	$7.21_{-0.01}^{+0.19}$	&	$7.40_{-0.02}^{+0.01}$	&	$7.40_{-0.02}^{+0.01}$	&	$7.40_{-0.20}^{+0.01}$	\\\\
&	Sigma (eV)	&	$1_{-0}^{+20}$	&	$1_{-0}^{+20}$	&	$1_{-0}^{+20}$	&	$1_{-0}^{+20}$	&	$1_{-0}^{+20}$	\\\\
&	Eqw (eV)	&	$250 \pm 110$	&	$260_{-230}^{+180}$	&	$270_{-10}^{+20}$	&	$300\pm 160$	&	$70_{-70}^{+400}$	\\\\

Co K$\beta$ / \ion{Ni}{xxvi}&	$I$ ($\times 10^{-6}$  ph s$^{-1}$ cm$^{-2}$)	&	$1.2 \pm 0.4$	&	$2.2 \pm 0.9$	&	$1.0 \pm 0.5 $	&	$0.8 \pm 0.5$	&	$0.1 \pm 0.1$	\\\\
&	Center (keV)	&	$7.7 \pm 0.1$	&	$7.5 \pm 0.1$	&	$7.7 \pm 0.1$	&	$7.7 \pm 0.1$	&	$7.7_{-0.2}^{+0.1}$	\\\\
&	Sigma (eV)	&	$20_{-20}^{+1}$	&	$20_{-20}^{+1}$	&	$1_{-0}^{+20}$	&	$1_{-0}^{+20}$	&	$1_{-0}^{+20}$\\\\
&	Eqw (eV)	&	$360\pm 120$	&	$440_{-170}^{+180}$	&	$350\pm 160$	&	$280\pm 160$	&	$400_{-400}^{+600}$	\\\\

\bottomrule									
\end{tabular}									
\end{adjustbox}									
\label{}									
\end{table*}

\section{Eclipsing systems parameters}

\begin{table*}												
\centering												
\begin{adjustbox}{max width=\textwidth}												
\begin{tabular}{lcccccccccc}												
\toprule

Source	&	Donor spectral type	&	Companion radius	&	Inclination $i$	&	Semi mayor axis $a$	&	Eccentricity $e$	&	$\omega$	&	$v_{\infty}$	&	$\dot{M}$	&	Ratio	&	$\delta$	\\\\
	&		&$R_{\odot}$		&	deg	& $R_{\odot}$	&		&	deg	&km s$^{-1}$		& $\times 10^{7}$ $M_{\odot}$ yr$^{-1}$	& deg	&$R_{\odot}$		\\\\

\hline

Cen X-3	&	O6,5 II-III (1)	&	$12.1 \pm 0.5$ (1)	&	$65 \pm 1$ (6)	&	$18.84 \pm 0.02$ (11)	& $\leqslant 0.0016$ (16)	&	$0 \pm 0$(6)	&	$2050 \pm 600$	(6)&	5.3(6)	&	9.79 (19) 	&	$16 \pm 1$	\\\\
LMC X-4	&	O8 III (1)	&	$7.8 \pm 0.4$  (1)	&	$59.3 \pm 9$ (6)		&	$13.19 \pm 0.04$ (12)	&	$0.0006 \pm 0.0002$ (12)	&	$0 \pm 0$(6)	&	$1950 \pm 600$(6)	&	2.4	(6)&	237.18* (19) 	&	$13 \pm 3$	\\\\
SMC X-1	&	B0 I (2)	&	$18 \pm $ (7)	&	$62 \pm 2$ (6)		&	$26.08 \pm 0$ (13)	&	$\leqslant 0.0007$ (7)	&	$166 \pm 12$(6)	&	$870 \pm 260$(6)	&	15(6)	&	76.62 (19) 	&	$24 \pm 1$	\\\\
4U 1700-377	&	O7f (3)	&	$21.9 \pm 1.3$ (8)	&	$62 \pm 1$ (6)		&	$31.69 \pm 4.2$(14)	&	$\leqslant 0.0008 $ (17)	&	$49 \pm 11$	(6)&	$1850 \pm 550$(6)	&	21(6)	&	8.01 (19) 	&	$30 \pm 1$	\\\\
4U 15382-22	&	B0Iab (4)	&	$17.2 \pm 1$ (9)	&	$67 \pm 1$ (6)		&	$24.84 \pm 37$ (15)	&	$0.174 \pm 0.015$(15)	&	$40 \pm 12$(6)	&	$1000 \pm 300$	(6)&	8.3	(6)&	30.28 (19) 	&	$19 \pm 1$	\\\\
IGR J180272-2016	&	B1-Ib (5)	&	$19.2 \pm 4.2$ (10)	&	$72 \pm 2$	 (6)	&	$30.78 \pm 0$ (10)	&	$0.2 \pm 0$	(18)&	$0 \pm 0$(6)	&	$680 \pm 200$(6)	&	6.3	(6)&	43.38 (19) 	&	$19 \pm 2$	\\\\
XTE J1855-026	&	B0Iaep (6)	&	$22 \pm 2$ (6)	&	$71 \pm 2$ (6)		& 	$36.65 \pm 17.27$ (6)	&	$0.04 \pm 2$ (6)	&	$226 \pm 15$	(6)&	$620 \pm 190$(6)	&	65 (6)	&	$\sim$ 70	&	$24 \pm 2$	\\\\

\bottomrule												
\end{tabular}												
\end{adjustbox}	
\caption{Eclipsing systems. References: (1) \citet{2007A&A...473..523V}, (2) \citet{10.1093/mnras/261.2.337}, (3) \citet{10.1093/mnras/163.1.7P}, (4) \citet{10.1093/mnras/184.1.73P}, (5) \citet{2010A&A...510A..61T}, (6) \citet{falanga}, (7) \citet{1977ApJ...217..543P}, (8), \citet{2002A&A...392..909C}, (9) \citet{10.1093/mnras/256.4.631}, (10) \citet{2005A&A...439..255H}, (11) \citet{10.1111/j.1365-2966.2009.15778.x}, (12) \citet{Levine_2000}, (13) \citet{1993ApJ...410..328L}, (14) \citet{1973ApJ...181L..43J}, (15) \citet{mukherjee2007orbital}, (16) \citet{Bildsten_1997}, (17) \citet{10.1093/mnras/stw1299}, (18) \citet{Augello_2003}, (19) \citet{2019ApJS..243...29A}. (* Mean value).}	
\label{ratio}												
\end{table*}


\bsp	
\label{lastpage}
\end{document}